\documentclass{article}

\PassOptionsToPackage{numbers, compress}{natbib}
\usepackage[final]{neurips_2020_ml4ps}

\usepackage[utf8]{inputenc} % allow utf-8 input
\usepackage[T1]{fontenc}    % use 8-bit T1 fonts
\usepackage{hyperref}       % hyperlinks
\usepackage{url}            % simple URL typesetting
\usepackage{amsfonts}       % blackboard math symbols
\usepackage{nicefrac}       % compact symbols for 1/2, etc.

\usepackage{caption}
\usepackage{subcaption}
\usepackage{physics}
\usepackage{graphicx}
\usepackage{amsfonts}
\usepackage{amsmath}
\usepackage{amssymb}

\title{Generative models for sampling of lattice field theories}

\author{%
    Matija Medvidović\\
    Center for Computational Quantum Physics, Flatiron Institute, New York, NY 10010, USA\\
    Department of Physics, Columbia University, New York 10027, USA\\
    \texttt{matija.medvidovic@columbia.edu}\\
    \And
    Juan Carrasquilla\\
    Vector Institute for Artificial Intelligence, MaRS Centre, Toronto, Ontario, Canada\\
    Department of Physics and Astronomy, University of Waterloo, Ontario, N2L 3G1, Canada\\
    \texttt{carrasqu@vectorinstitute.ai}\\
    \And
    Lauren E. Hayward\\
    Perimeter Institute for Theoretical Physics, Waterloo, Ontario, N2L 2Y5, Canada\\
    \texttt{lhayward@perimeterinstitute.ca}\\
    \And
    Bohdan Kulchytskyy\\
    Department of Physics and Astronomy, University of Waterloo, Ontario, N2L 3G1, Canada\\
    Perimeter Institute for Theoretical Physics, Waterloo, Ontario, N2L 2Y5, Canada\\
    \texttt{bkulchytskyy@uwaterloo.ca}\\
}

\begin{document}

\maketitle

\begin{abstract}
We explore a self-learning Markov chain Monte Carlo method based on the Adversarial Non-linear Independent Components Estimation Monte Carlo, which utilizes generative models and  artificial neural networks. 
We apply this method to the scalar $\varphi^4$ lattice field theory in the weak-coupling regime and, in doing so, greatly increase the system sizes explored to date with this self-learning technique. 
Our approach does not rely on a pre-existing training set of samples, as the agent systematically improves its performance by bootstrapping samples collected by the model itself. 
We evaluate the performance of the trained model by examining its mixing time and study the ergodicity of generated samples. When compared to methods such as Hamiltonian Monte Carlo, this approach provides unique advantages such as the speed of inference and a compressed representation of Monte Carlo proposals for potential use in downstream tasks. 
\end{abstract}

%%%%%%%%%%%%%%%%%% INTRODUCTION %%%%%%%%%%%%%%%%%%
\section{Introduction}

Field theory is the theoretical bedrock for unified 
descriptions of critical phenomena. 
Such a framework provides us with a microscopic understanding of universality~\cite{cardy_1996}. 
Its predictions for a variety of strongly correlated systems have been confirmed experimentally to high precision in widely diverse physical systems~\cite{sachdev2001}, including thin films of superfluids~\cite{RevModPhys.80.1009}, superconductors~\cite{Halperin1979}, ferromangets~\cite{coldea2010a}, quantum simulations of frustrated magnets~\cite{king2018}, and ultracold atoms~\cite{greiner2002}. 
In practice, however, field theories are often intractable and their analytical treatment involves approximations with varying degrees of reliability. Luckily, some of these field theories lend themselves to non-perturbative numerical simulations. These simulations often utilize Markov Chain Monte Carlo (MCMC) algorithms based on a Feynman path-integral formulation of the theory~\cite{RevModPhys.20.367}, tensor networks~\cite{PhysRevLett.110.100402, PhysRevLett.118.220402, PhysRevLett.104.190405, Magnifico2020}, or even quantum computing~\cite{preskill2018}.

A key practical concern in MCMC simulations is the autocorrelation that exists between Monte Carlo samples~\cite{Newman1999}. Reducing the autocorrelation time enables a Markov chain to become shorter while maintaining the same statistical predictive capacity. Such optimization can be achieved through a tailored design of the proposal distribution in an MCMC update. The best proposals dramatically reduce the autocorrelation and thus the computation time required to reach a desired accuracy. 

Recently, there has been progress toward an automatic optimization of the proposal distribution in MCMC simulations applied to physical systems. 
Specifically, the proposal distribution is parameterized as a generative model designed for an inexpensive generation of statistically independent samples, such as a generative adversarial network (GAN) \cite{goodfellow2014, Arjovsky2017} or a flow model \cite{Dinh2014, Dinh2016, Kingma2018_2}. 
An important ingredient in such optimization is the choice of a loss function that ultimately leads to an efficient sampler. 
For example, Ref.~\cite{Urban2018} takes a data-driven approach relying on a GAN's loss-function evaluated over a pre-existing data-set of samples. 
Contrarily, Refs.~\cite{Albergo2019, Boyda2020, Kanwar2020, Nicoli2020} take a variational approach that aims to minimize the action for the samples generated from the model.

Here we examine a general approach for a self-training MCMC known as Adversarial Nonlinear Independent Components Estimation Monte Carlo (A-NICE MC) \cite{Song2017}, where a neural network is optimized to minimize the autocorrelation in a Markov chain. The network is trained on samples generated by the model itself and, like most MCMC schemes, requires only the analytical lattice action as input. By extending the A-NICE MC method to sample high-dimensional lattice field theories, we find that this method scales well beyond previously tested target space dimensionalities. 
Using A-NICE MC on the scalar field $\varphi ^4$ theory \cite{Weinberg1995} proves that the model can be systematically optimized towards producing decorrelated samples using standard gradient-based optimizers. The $\varphi ^4$ theory was chosen because of its simplicity.
In addition, we study the performance of A-NICE MC by examining its ability to discriminate between random Gaussian noise and actual samples from the weakly coupled theory. 
We find that the model learns to distinguish between the two and that we can observe the distinction in the training procedure. 
While the training procedure for A-NICE MC is considerably more expensive than Hamiltonian Monte Carlo (HMC), the strategy has the advantage that a trained A-NICE MC model can simply be stored, shared, and sampled when needed, as opposed to HMC where the result of the calculations is a collection of samples. 

%%%%%%%%%%%%%%%%%% METHODS %%%%%%%%%%%%%%%%%%
\section{Methods}

We focus on obtaining lattice field samples $\{ \varphi ^{(i)} \}$ from a Boltzmann-type distribution of the form $P (\varphi) \propto \exp (-S (\varphi))$. The distribution is defined on a square $L \times L$ lattice $\Lambda = \mathbb{R} ^{L^2}$, where each sample $\varphi \in \Lambda$. Metropolis-Hastings (MH)~\cite{Metropolis1953,Hastings1970} MCMC algorithms propose each new sample $\varphi ^{(i+1)}$ from $\varphi ^{(i)}$ through a predefined proposal function $f$. Each $\varphi ^{(i+1)}$ is either accepted and appended to the Markov Chain, or rejected such that $\varphi ^{(i)}$ is appended to the chain instead. Both detailed balance and ergodicity are required of the mapping $\varphi ^{(i)} \mapsto \varphi ^{(i+1)}$. The details of the MH algorithms can be found in Ref.~\cite{Newman1999}. Ideally, this probabilistic mapping is chosen such that the statistical correlation between adjacent samples is minimized.

An important and widely used algorithm in the study of lattice quantum field theory is the HMC algorithm~\cite{Neal2012}. This method supplements the original degrees of freedom $\varphi$ with fictitious conjugate momentum variables $\pi$ and introduces a combined Hamiltonian on the same lattice $\Lambda$ defined as $H _\Lambda (\varphi ,\pi ) = \frac{1}{2}\sum _{\vb{x} \in \Lambda} \pi _{\vb{x}} ^2 + S _\Lambda (\varphi )$. 
A state proposal $(\varphi, \pi) ^{(i+1)}$ is then obtained by integrating Hamilton's equations for a finite time. 
Upon integrating out the momentum variables in the distribution $P _{H_\Lambda} (\varphi, \pi) \propto \exp (- H _\Lambda)$, one recovers samples from the desired $P (\varphi)$. 
Since the Hamiltonian dynamics generate a volume-preserving and time-reversible flow of states, the proposal distribution is symmetric, implying that the rate of rejection of proposed samples is determined by the accuracy of the numerical integrator ~\cite{Neal2012}.

A-NICE MC takes inspiration from HMC and embeds its proposal distribution in the augmented space $f:(\varphi, \pi) ^{(i)}\rightarrow (\varphi,\pi) ^{(i+1)} $ where the supplementary degrees of freedom $\pi$ are Gaussian distributed. Furthermore, A-NICE exploits the invertible and volume-preserving properties of the  NICE~\cite{Dinh2014} architecture to parametrize $f$. As a result, the proposal distribution is symmetrical and does not rely on expensive integration as in HMC.

The NICE proposal $f_\theta$, which is parametrized by $\theta$, is trained alongside an adversary to systematically minimize autocorrelation times. This setup closely resembles a more traditional GAN~\cite{goodfellow2014, Arjovsky2017, Chen2016} but retains a few key differences. Most notably, a pairwise discriminator network is used in place of a cost function. 
Instead of ranking each sample individually as for a more conventional discriminator, the pairwise discriminator ranks pairs of samples together, acting as a proxy for the auto-correlation function between samples. The discriminator is parametrized as a deep neural network and represents a mapping $D_\alpha : \mathbb{R} ^{2n} \rightarrow \mathbb{R}$ with parameters $\alpha$. Intuitively, the discriminator takes a pair of samples and is trained to score the degree of their correlation. The objective is to produce a NICE proposal that generates a Markov chain with two desirable properties.
Firstly, starting from noise, the proposal should generate a highly-probable sample in a small number of steps such that equilibrium is reached quickly.
Secondly, starting from a highly-probable sample, the proposal should generate another decorrelated and probable sample in as few steps as possible.

We employ a bootstrapping procedure to generate a training dataset. 
The untrained network is initially sampled to obtain a starting dataset $\mathcal{D} _0$, which is expected to exhibit large autocorrelation times. Still, the MH algorithm biases the sampling towards the correct distribution regardless of the proposal \cite{Newman1999}. 
We collect samples from multiple chains and shuffle them to further reduce hidden correlations in the training set. 
After $N_\text{bootstrap}$ training loop iterations using $\mathcal{D} _0$, some fraction $r \in [0,1]$ of samples $\mathcal{D} _0$ are replaced by new data, yielding $\mathcal{D} _1$. 
This new dataset is less correlated since it comes from the partially trained model. 
Iterating this procedure, we obtain samples of increasing quality in each subsequent dataset $\mathcal{D} _i$. In principle, we reach the fixed point of this training procedure when the samples generated by $f _\theta$ become as decorrelated as the shuffled samples from the bootstrapped dataset $\mathcal{D} _i$.

The scalar $\varphi ^4$ theory on a lattice was chosen as the simplest non-Gaussian theory to test the effectiveness of the A-NICE MC approach. Despite its simplicity, this theory has a nontrivial phase diagram and a variety of use cases across particle and many-body physics. As noted earlier, in this case, samples consist of only one real number per lattice site. For the $\varphi^4$ lattice field theory, the probability distribution $P _\Lambda (\varphi ) \propto e^{-S_\Lambda (\varphi )}$ is defined by the action
\begin{equation}
    S _\Lambda (\varphi ) = \sum _{\vb{x} \in \Lambda} \left[ -2\kappa \sum _{\mu = 1} ^D  \varphi _{\vb{x}} \, \varphi _{\vb{x} + \hat{\vb{e}} _\mu}
    + (1-2\lambda)\; \varphi _{\vb{x}} ^2 + \lambda \, \varphi _{\vb{x}} ^4 \right],
\label{phi4_action}
\end{equation}
where the coupling constants $\kappa, \gamma \in \mathbb{R}$ define different distributions $P _\Lambda (\varphi )$ and determine the phase diagram of the model. 
We choose $\kappa = 0.21$ and $\lambda = 0.022$, for which the model is in the weak-coupling regime and displays behavior similar to the paramagnetic Ising model~\cite{Kardar2012}. The phase diagram of the theory associated with Eq.~\ref{phi4_action} was explored in Ref.~\cite{De2005} using standard MCMC algorithms. 
Our choice of parameters enables us to tackle our goal of exploring novel MCMC frameworks for a non-Gaussian model, while eliminating any strong-coupling effects.
Remaining in the weakly-coupled regime allows us to study A-NICE MC performance in the region of parameter space easily reachable by conventional methods which allows for a straightforward comparison of results.

%%%%%%%%%%%%%%%%%% RESULTS %%%%%%%%%%%%%%%%%%
\section{Results}

\begin{figure*}[t]
    \centering
    \includegraphics[width=0.9\textwidth]{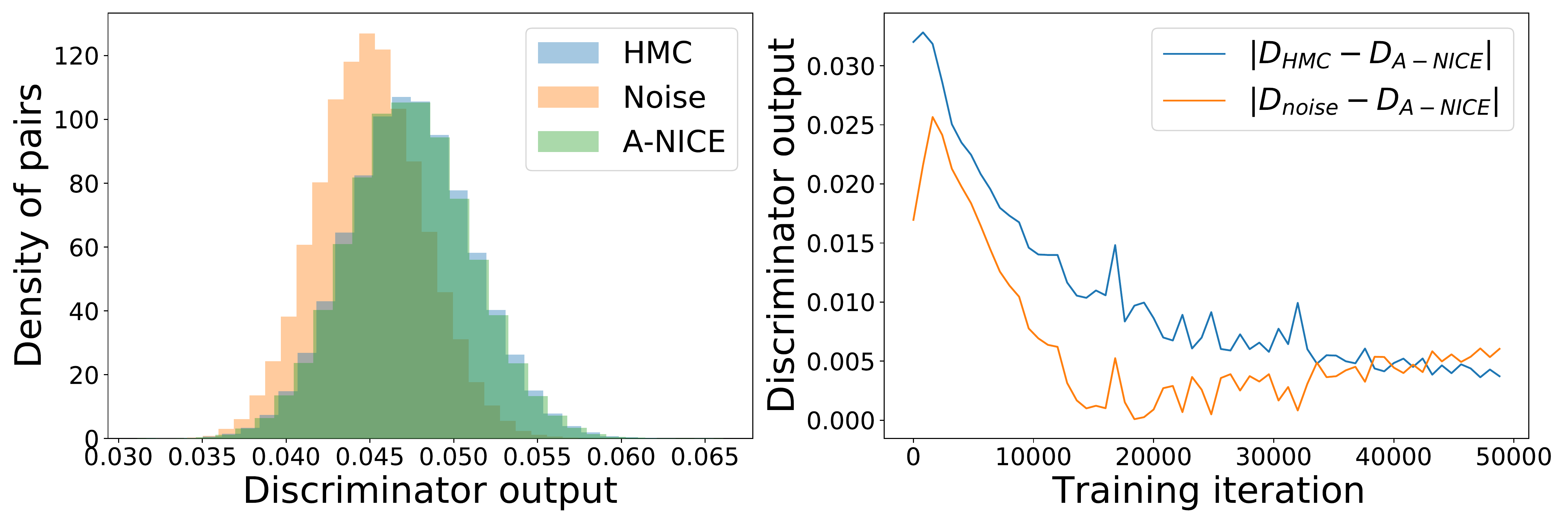}
    \caption{(left) The fully-trained discriminator score distribution of randomly shuffled HMC, A-NICE and (unit) Gaussian random noise for an $8 \times 8$ lattice. All three histograms have been normalized to unity. The noise was chosen with the mean and the variance estimated from HMC samples. (right) The mean value of the same three scores with respect to the training step and averaged over 32 independent chains.}
    \label{ergodicity}
\end{figure*}

It is difficult a priori to verify whether the A-NICE framework will display ergodic behavior since it is stochastically optimized, highly nonlinear, and contains a large number of free parameters. 
We thus compare our A-NICE algorithm against well-established HMC simulations. 
First, we take the generated HMC chain and randomly pair the samples (to decorrelate them artificially) in order to examine the output of the fully-trained A-NICE pairwise discriminator. The results are shown in Fig.~\ref{ergodicity}.
We observe that the A-NICE and HMC output distributions overlap, which means that the A-NICE distribution is consistent with the behavior of the HMC distribution in that the produced samples are not penalized differently by the discriminator and pushed outside the bulk of the distribution. The fact that all HMC samples being scored similarly suggests that the discriminator had seen all such samples from the NICE proposal during training.
We supplement the plot with a reference point provided by samples randomly chosen from a unit Gaussian distribution. 
Since we are studying our lattice model well within the weakly coupled regime, we expect our field samples to be almost Gaussian and thus penalized in a similar way.
We plot the discriminator scores of HMC, A-NICE and the Gaussian random noise against the training iteration in the right panel of Fig.~\ref{ergodicity}. 
Differences between discriminator scores of A-NICE samples and HMC/random samples are shown, where the HMC samples are randomly shuffled to artificially eliminate autocorrelations and provide reliable reference values. 
We see that, as the discriminator optimizes \cite{Arjovsky2017}, the difference $\vert D_{\text{HMC}} - D_{\text{A-NICE}}\vert$ decreases and thus the A-NICE score approaches the decorrelated HMC score.
We perform error analysis on the fully-trained distributions in the left panel of Fig.~\ref{ergodicity} and find that $\vert D_{\text{HMC}} - D_{\text{A-NICE}} \vert = (416.7 \pm 1.4) \times 10^{-5}$, which is less than $\vert D_{\text{noise}} - D_{\text{A-NICE}} \vert = (433.0 \pm 1.5) \times 10^{-5}$. Therefore, we conclude that A-NICE MC learns to distinguish between random noise and a weakly-coupled $\varphi ^4$ theory, since the A-NICE samples become less similar to random noise and become more similar to decorrelated HMC samples by the end of the training.

\begin{figure*}[t]
    \centering
    \includegraphics[width=0.7\textwidth]{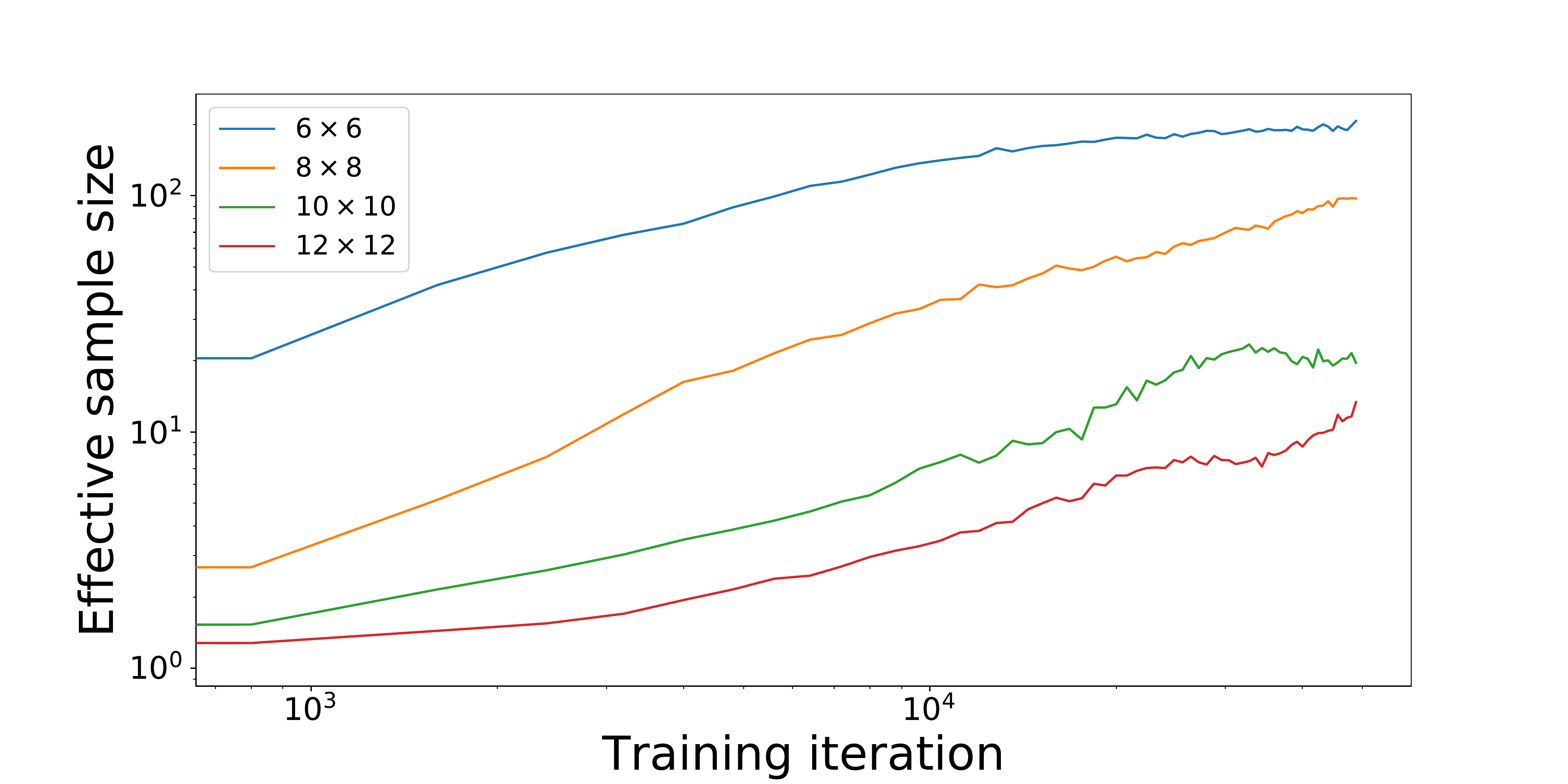}
    \caption{Effective sample size (ESS) as a function of training iteration. The increasing ESS demonstrates A-NICE MC's ability to systematically improve the statistical quality metrics of the Markov chain produced by the model.}
    \label{training_ess}
\end{figure*}

In Fig.~\ref{training_ess}, we study the systematic improvement of sample quality during the training process.
Specifically, we plot the effective sample size (ESS), which is inversely proportional to the integrated autocorrelation time $\tau$ \cite{Newman1999} of the chain (ESS $=N/2\tau$, where $N$ is the number of samples) and approximately represents the effective number of independent samples in a given chain.
We expect that additional time and computational resources would continue to improve the model.

\begin{figure*}[t]
    \centering
    \includegraphics[width=0.95\textwidth]{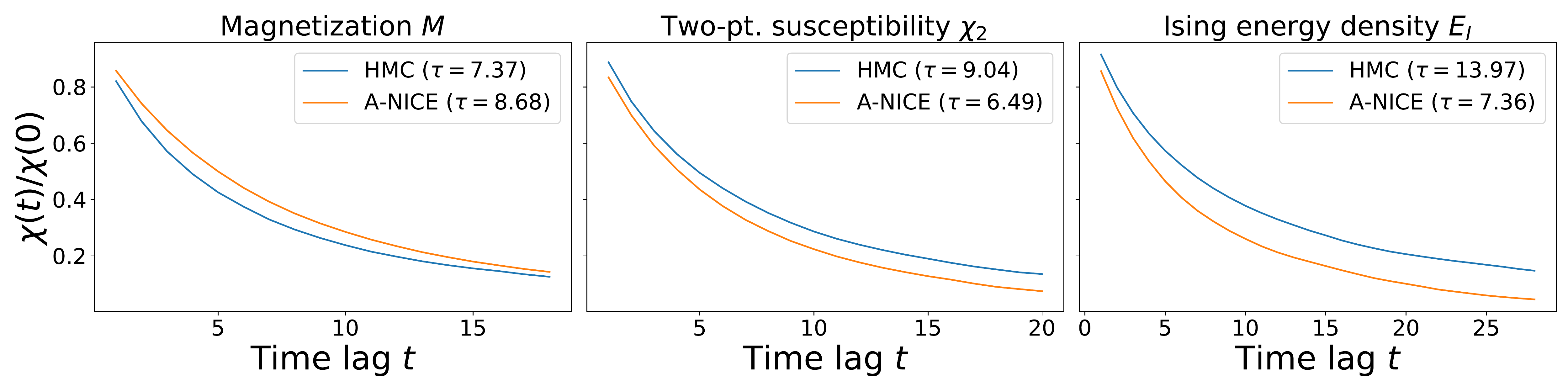}
    \caption{Comparisons of common physical observables  
    as a function of the time difference (lag) $t$ between two MCMC samples for $L=10$. 
    Each $y$-axis shows the normalized statistical autocorrelation $\chi(t) = \int _0 ^\infty \dd t' \; \left[ \mathcal{O}(t') \mathcal{O} (t + t') - \langle \mathcal{O} \rangle ^2 \right]$ for a different observable $\mathcal{O}$.
    Integrated autocorrelation times \cite{Newman1999} for both methods and each observable are reported in the figure legends.
    }
    \label{ac_decay}
\end{figure*}

Next, we study the behaviour of some observables, including the magnetization 
$M \equiv \left\langle \frac{1}{L^D} \sum _{\vb{x} \in \Lambda} \varphi _{\vb{x}} \right\rangle$, 
the two point-susceptibility 
$\chi _2 \equiv \sum _{\vb{x} \in \Lambda} G _2 (\vb{x})$ (where 
$G _2 (\vb{x}) \equiv \frac{1}{L^D} \sum _{\vb{y} \in \Lambda} \left[ \left\langle \varphi _{\vb{y}} \; \varphi _{\vb{x} + \vb{y}} \right\rangle - \left\langle \varphi _{\vb{y}} \right\rangle \left\langle \varphi _{\vb{x} + \vb{y}} \right\rangle  \right]$), 
and the Ising energy density $E _I \equiv \lim _{\lambda \rightarrow \infty} \frac{1}{D} \sum _\mu G _2 (\hat{\vb{e}} _\mu )$.
As illustrated in Fig.~\ref{ac_decay}, A-NICE outperforms HMC for most observables of interest since the A-NICE autocorrelation curves decay faster, except for the case of the average magnetization. Note that HMC can be tuned to improve performance, but usually at the cost of increased computational complexity. Since HMC constructs proposals by integrating differential equations, more steps would be required each time. Thus, the proposal would be further away from the current state, lowering autocorrelations and decreasing acceptance rates. A-NICE doesn't force such a trade-off but offers no control over acceptance rates in return.
For HMC, we take 10 leapfrog steps for each proposal and the step size is tuned so that the acceptance rate matches A-NICE MC.

%%%%%%%%%%%%%%%%%% CONCLUSION %%%%%%%%%%%%%%%%%%
\section{Conclusion}

We have applied the A-NICE MC \cite{Song2017} algorithm to the lattice $\varphi ^4$ field theory in the weakly-coupled regime, increasing the number of dimensions that have been successfully sampled using this method to date. 
The model can be systematically optimized and produces samples which are consistent with an ergodic exploration of the state-space. The bulk of the computational cost for employing this technique is associated with the training phase. Deployment of the trained model is relatively inexpensive compared to the HMC method tuned to the same performance level. This sampling efficiency of the trained model effectively eliminates the need to store the samples, which cuts down on the storage space requirements and allows us to extract physical observables with low statistical uncertainties. 

In future work, it will be interesting to extend our experiments with A-NICE MC to the strongly-interacting regime of the $\varphi ^4$ field theory as well as to apply this framework to more complicated theories with additional internal degrees of freedom or gauge symmetries. We conjecture that A-NICE MC can be extended to deal with these models with possible modifications required to efficiently explore the state space in a systematic domain-specific way. 
\section*{Impact statement}

Our Adversarial Nonlinear Independent Components Estimation Monte Carlo (A-NICE MC) approach to sampling focuses on training a neural network to be embedded into Markov chain Monte Carlo (MCMC) sampling algorithms. A-NICE MC moves most of the computational cost of MCMC sampling to network training, making sampling itself relatively less expensive compared to established MCMC approaches.

We expect that further improvements to the adversarial network architecture are possible such that the sampling can continue to become even more efficient in the future.
Improved efficiency in MCMC sampling schemes has the potential to greatly reduce the overall carbon footprint of high-performance computational (HPC) science. Lower autocorrelation times mean fewer sampling steps which, in turn, mean more effective samples per watt of consumed power. Therefore, we expect that, given further optimizations, machine-learning-based sampling methods such as A-NICE MC have the potential to significantly reduce the scientific carbon footprint. 

\bibliographystyle{utphys}
\bibliography{References}

\end{document}